\documentclass[10pt]{iopart}
\usepackage{graphicx,color}

\begin{document}

\title{On the conservation of helicity in a chiral medium}

\author{Frances Crimin$^1$, Neel Mackinnon$^{1}$, J\"{o}rg B. G\"{o}tte$^{1,2}$ and Stephen M. Barnett$^1$}

\address{$^1$Department of Physics and Astronomy, University of Glasgow,
Glasgow G12 8QQ, United Kingdom

{$^2$College of Engineering and Applied Science, Nanjing University, Nanjing 210093, China}}

\ead{frances.crimin@glasgow.ac.uk}
\vspace{10pt}
\begin{indented}
\item[]July 2019
\end{indented}

\begin{abstract}
We consider the  energy and helicity densities of circularly polarised light within a lossless chiral medium, characterised by the chirality parameter $\beta$. A form for the helicity density is introduced, valid to first order in $\beta$, that produces  a helicity  of $\pm\hbar$ per photon for right and left circular polarisation, respectively. This is in contrast to the result obtained if we use the form of the helicity density employed for linear media.  We examine the helicity continuity  equation, and show that this modified form of the helicity density  is required  for consistency with  the dual symmetry condition of a chiral medium with a constant value of $\epsilon/\mu$.   Extending the results to  arbitrary order in $\beta$ establishes an exact relationship between the energy and helicity densities in a chiral medium.
\end{abstract}

\maketitle
\ioptwocol

\section{Introduction}

The helicity of an electromagnetic field continues to receive interest as a way of describing the interaction of chiral light with matter~\cite{Barnett2012,Fernandez-Corbaton2013a,VanKruining2016,Nienhuis2016,Alpeggiani2018}. 
Interchanging the  electric and magnetic fields in the free-space Maxwell equations leaves them invariant, as a consequence of their dual symmetry~\cite{Jackson1962}. From Noether's theorem~\cite{Calkin1965,Cameron2012,Bliokh2013}, the conserved quantity arising from this symmetry is indeed the electromagnetic helicity of the fields, which characterises their twist, or vorticity. In the presence of matter, helicity is no longer generally conserved, but one can still write down a continuity equation which shows how currents and charges can act as sources or sinks of helicity, analogous to the continuity equation for electromagnetic energy~\cite{Nienhuis2016}.  Thus the (non)-conservation of the  helicity of a field can be used to characterise  different types of matter~\cite{Crimin2019}.

The study of electromagnetic helicity within media, as opposed to in the free electromagnetic field,  has been undertaken in recent years~\cite{Fernandez-Corbaton2013a,VanKruining2016,Alpeggiani2018}. The conditions under which helicity is conserved in a lossless linear, isotropic  medium were considered by Fernandez-Corbaton \emph{et al.}~\cite{Fernandez-Corbaton2013a}, with the results extended to include anisotropic media by van Kruining and G\"{o}tte in~\cite{VanKruining2016}.   The definition of helicity in dispersive, lossless media has been examined by Alpeggiani \emph{et al.}~\cite{Alpeggiani2018}, while the electromagnetic chirality, proportional to the helicity in the case of monochromatic fields, has been examined in lossy media by V\'azquez-Lozano and Mart\'inez~\cite{Vazquez-Lozano2018}.

In this paper, we discuss helicity in a  {dual-symmetric, homogeneous, isotropic and lossless chiral medium. We use the conservation of energy and helicity in such media to determine an appropriate expression for the helicity density. We first examine both the energy denisty and helicity density   to first order in the chirality parameter, $\beta$, which itself describes the chiroptical response of the medium.} {From this result, it follows that retaining terms of $\mathcal{O}(\beta)$ in the energy density is necessary for conservation of the helicity density to the same order.} Such a chiral contribution  to the helicity density is required both to satisfy the dual symmetry condition, and to produce a helicity of {$\pm\hbar$} per photon for right- and left-handed circular polarised light within the medium.  We extend the results to examine both the energy and helicity density to arbitrary  orders in $\beta$ and propose an exact relationship between the two, showing that  this  relationship is a direct consequence of their conservation to all orders in the chirality parameter. {The results in this paper have been stated in summary in our recent review on helicity and chirality~\cite{Crimin2019}. The present work provides a more complete analysis and derivation of our results. }

\section{Helicity density in a chiral medium}

\subsection{Dual symmetry and helicity conservation}

Electric-magnetic ``democracy'' ~\cite{Berry2009} in the absence of charge is perhaps most striking when we express both electric and magnetic fields in terms of the vector potentials $\mathbf{A}$ and $\mathbf{C}$~\cite{Stratton1941,Cameron2014a}:
\begin{eqnarray}
\mathbf{B}=\mbox{\boldmath$\nabla$} \times\mathbf{A},\;\;\;\;\mathbf{D}=-\mbox{\boldmath$\nabla$} \times\mathbf{C}.
\end{eqnarray}
If we chose a gauge such that $\mbox{\boldmath$\nabla$}  \cdot \mathbf{A} = 0$, and also $\mbox{\boldmath$\nabla$}  \cdot \mathbf{C} = 0$, then using Maxwell's equations allows us to relate the fields to the time derivatives of the potentials:
\begin{eqnarray}
\mathbf{E}=-\dot{\mathbf{A}},\;\;\;\;\mathbf{H}=-\dot{\mathbf{C}}.
\end{eqnarray}

In the presence of matter, the symmetry between electric and magnetic fields no longer holds, as matter is comprised only of electric charges with no magnetic ones. {We should note, however, that fields very much like those of a magnetic monopole can emerge as a result of many-body interactions in spin ice~\cite{Castelnovo2008}}.   In some circumstances, the idea of electric-magnetic democracy can be generalised to hold even in media, provided the effects of the charges comprising the medium are treated using macroscopic electrodynamics. To demonstrate this, substituting  the Drude-Born-Fedorov (DBF) constitutive relations for a chiral medium~\cite{Lakhtakia1994}
\begin{eqnarray}\label{eq:DBF}
&\mathbf{D}=\epsilon\left(\mathbf{E}+\beta\mbox{\boldmath$\nabla$} \times\mathbf{E}\right),\nonumber\\
&\mathbf{B}=\mu\left(\mathbf{H}+\beta\mbox{\boldmath$\nabla$} \times\mathbf{H}\right),
\end{eqnarray}
into Maxwell's equations, and performing the duality transformation~\cite{Jackson1962}
\begin{eqnarray}
\mathbf{E}\rightarrow\mathbf{E}\cos\theta+\sqrt{\frac{\mu}{\epsilon}}\mathbf{H}\sin\theta,\nonumber\\
\mathbf{H}\rightarrow\mathbf{H}\cos\theta-\sqrt{\frac{\epsilon}{\mu}}\mathbf{E}\sin\theta,\nonumber\\
\mathbf{D}\rightarrow\mathbf{D}\cos\theta+\sqrt{\frac{\epsilon}{\mu}}\mathbf{B}\sin\theta,\nonumber\\
\mathbf{B}\rightarrow\mathbf{B}\cos\theta-\sqrt{\frac{\mu}{\epsilon}}\mathbf{D}\sin\theta,
\end{eqnarray}
leaves the equations  invariant if $\mbox{\boldmath$\nabla$} (\epsilon/\mu)=0.$ In other words, the condition for duality symmetry within a linear medium is that the ratio $\epsilon/\mu$ remains constant~\cite{Fernandez-Corbaton2013a,VanKruining2016}. Conservation of helicity is associated with dual symmetry~\cite{Barnett2012,Fernandez-Corbaton2013a}, so this is also the condition under which helicity is conserved. These considerations {leave  the} chirality parameter $\beta$ unspecified, and it can vary freely in space without interfering with the dual symmetry of the system~\cite{VanKruining2016}.

Returning to free space, we can use the free-field Maxwell equations to write down a continuity equation for the helicity~\cite{Barnett2012,Nienhuis2016}
\begin{eqnarray}\label{eq:continuity}
\partial_t h+\mbox{\boldmath$\nabla$} \cdot\mathbf{v}=0, 
\end{eqnarray}
where 
\begin{eqnarray}\label{eq:helicityvacuum}
h=\frac{1}{2}\left(\sqrt{\frac{\epsilon_0}{\mu_0}}\mathbf{A}\cdot\mathbf{B}-\sqrt{\frac{\mu_0}{\epsilon_0}}\mathbf{C}\cdot\mathbf{D}\right)
\end{eqnarray}
is the  helicity density of a free electromagnetic field,  with  the associated flux density $\mathbf{v}$  given by 
\begin{eqnarray}\label{eq:fluxvacuum}
\mathbf{v}=\frac{1}{2}\left(\sqrt{\frac{\epsilon_0}{\mu_0}}\mathbf{E}\times\mathbf{A}+\sqrt{\frac{\mu_0}{\epsilon_0}}\mathbf{H}\times\mathbf{C}\right).
\end{eqnarray}
We recognise this as the dual-symmetric form of the spin density multiplied by the speed of light~\cite{Barnett2010a}.
The continuity equation for helicity can be compared with the continuity equation for energy $\partial_t w+\mbox{\boldmath$\nabla$} \cdot\mathbf{S}=0$~\cite{Cameron2017a}, where $
w=1/2\left(\epsilon_0|\mathbf{E}|^2+\mu_0|\mathbf{H}|^2\right)$
and $\mathbf{S}=\mathbf{E}\times\mathbf{H}$ 
are the familiar energy density and flux density of the free electromagnetic field. 

If we consider  right- and left-handed circularly polarised plane waves in vacuum with complex field components {
$
{\mathbf{E}^{\pm}_0}=E_0\exp\left[i(kz-wt)\right](\hat{\mathbf{x}}\pm i\hat{\mathbf{y}})$ and $
\mathbf{H}^{\pm}_0=\sqrt{\frac{\epsilon_0}{\mu_0}}E_0\exp\left[i(kz-wt)\right](\hat{\mathbf{y}}\mp i\hat{\mathbf{x}})$},
it is straightforward to show that the ratio of helicity density to energy density is{
\begin{eqnarray}\label{eq:densityratios}
\frac{\Re\left[\sqrt{\frac{\epsilon_0}{\mu_0}}\mathbf{A}^{\pm}_0\cdot(\mathbf{B}^{\pm}_0)^*-\sqrt{\frac{\mu_0}{\epsilon_0}}\mathbf{C}^{\pm}_0\cdot(\mathbf{D}^{\pm}_0)^*\right]}{\Re\left[\epsilon_0\mathbf{E}^{\pm}_0\cdot(\mathbf{E}^{\pm}_0)^*+\mu_0\mathbf{H}^{\pm}_0\cdot(\mathbf{H}^{\pm}_0)^*\right]}=\pm\frac{1}{\omega}
\end{eqnarray}}for the right and left handed polarisations. This is in accordance with the fact that the waves possess a helicity of $\pm\hbar$ per photon~\cite{Barnett2012}. Similarly, the ratio of flux densities along the direction of propagation  gives {
\begin{eqnarray}\label{eq:fluxratios}
\frac{\Re\left[\sqrt{\frac{\epsilon_0}{\mu_0}}\mathbf{E}^{\pm}_0\times(\mathbf{A}^{\pm}_0)^*+\sqrt{\frac{\mu_0}{\epsilon_0}}\mathbf{H}^{\pm}_0\times(\mathbf{C}^{\pm}_0)^*\right]\cdot \hat{\mathbf{z}}}{2\Re\left[\mathbf{E}^{\pm}_0\times(\mathbf{H}^{\pm}_0)^*\right] \cdot \hat{\mathbf{z}} }=\pm\frac{1}{\omega}.
\end{eqnarray}}We note that because the ratios of energy and helicity density are constant, and both the energy and helicity are locally conserved in vacuum, the helicity - like the energy - must travel at the speed of light. 

The definitions (\ref{eq:helicityvacuum}) and (\ref{eq:fluxvacuum}) can be extended to linear media with the replacements $\epsilon_0\rightarrow\epsilon$ and $\mu_0\rightarrow\mu$~\cite{VanKruining2016}. However, a simple example shows that the result (\ref{eq:densityratios}) does not follow from this extensions of (\ref{eq:helicityvacuum}) and (\ref{eq:fluxvacuum}) if the medium is chiral. This is because in a chiral medium, the energy density is not simply equal to $w=1/2\left(\epsilon|\mathbf{E}|^2+\mu|\mathbf{H}|^2\right)$, {but is instead given by~\cite{Barnett2016a,Bursian1926}
\begin{eqnarray}\label{eq:energydensity}
  w_{1}=\frac{1}{2}\left[\mathbf{D}\cdot\mathbf{E}+\mathbf{B}\cdot\mathbf{H}\mathbf{-}\beta\epsilon\mu(\mathbf{E}\cdot\dot{\mathbf{H}}-\dot{\mathbf{E}}\cdot\mathbf{H})\right],
\end{eqnarray}
where the subscript ``1" is used to indicate that this expression holds to $\mathcal{O}(\beta)$}.  Furthermore, it will be shown that {the straightforward extension of definition (\ref{eq:helicityvacuum}) by replacement of $\epsilon_0\rightarrow\epsilon$ and $\mu_0\rightarrow\mu$ for chiral media} is inconsistent with the requirement that helicity is conserved when $\mbox{\boldmath$\nabla$}(\epsilon/\mu)=0$.

We can motivate an expression for the helicity density in a chiral medium which preserves the results (\ref{eq:densityratios}) and (\ref{eq:fluxratios}) by considering a specific case in which we know that the helicity and energy densities must have the same relationship to one another as they do in vacuum. This is the case when $\mbox{\boldmath$\nabla$} (\epsilon/\mu) = 0$ throughout the space under consideration, including at interfaces,  as helicity and energy are then both locally conserved~\cite{Fernandez-Corbaton2013a,VanKruining2016}. We can then use our knowledge of the energy density to suggest an appropriate helicity density. 

\begin{figure}
\centering
\includegraphics[scale=0.4]{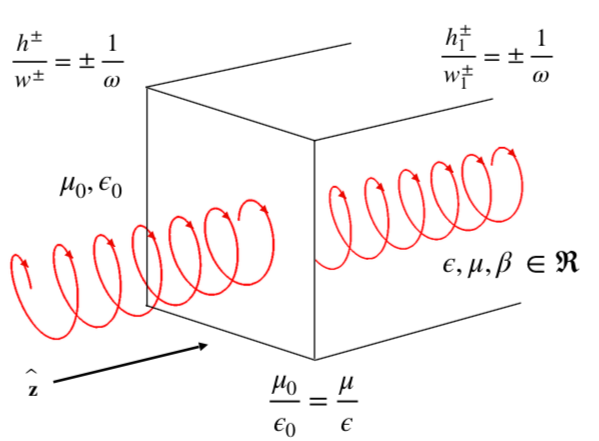}\caption{Consider light traversing the interface between vacuum and a dual-symmetric, lossless chiral medium characterised by $\epsilon$, $\mu$ and the chirality parameter $\beta$. As both the energy and helicity of the field are conserved, the ratio of the helicity density to energy density, $h/w$, is preserved across the interface. The energy density in a chiral medium, $w_1$ (\ref{eq:energydensity}), contains a term proportional to  $\beta$, such that the expression for the helicity density must be modified to $h_1$ (\ref{eq:helicitybeta}) in order that the ratio $h/w=h_1/w_1$ is maintained.}
\label{fig:1} 
\end{figure}

With this in mind, we consider the propagation of light from vacuum into a lossless, homogeneous and isotropic chiral medium, characterised by the constitutive relations (\ref{eq:DBF}). We also imagine that the chiral medium possesses permittivity and permeability such that $\epsilon / \mu = \epsilon_0/\mu_0$, which means helicity is conserved at the interface. This set-up is depicted in Figure (\ref{fig:1}). As there are neither sources nor sinks of helicity or energy, we expect the flux densities of these two quantities to remain unchanged from their vacuum values, and therefore that their ratio also remains unchanged from (\ref{eq:fluxratios}). We further expect that both the energy and helicity density  travel at the group velocity of the wave inside the medium, from which it follows that the ratio of the helicity density to energy density (\ref{eq:densityratios})  must be conserved across the dual-symmetric interface. We  use this to  postulate a helicity density, $h_1$, with this property. 
\subsection{The ratio of energy and helicity in a chiral medium}

First, we consider the helicity and energy fluxes. The electric and magnetic fields of right- and left-handed circularly polarised plane waves in the chiral medium can be written as~\cite{Lekner1996}
\begin{eqnarray}\label{eq:EandH}
\mathbf{E}^{\pm}=E_{0}\exp\left[i(kz-\omega t)\right](\hat{\mathbf{x}}\pm i\hat{\mathbf{y}}),\\
\mathbf{H}^{\pm}=\sqrt{\frac{\epsilon}{\mu}}E_0\exp\left[i(kz-\omega t)\right](\hat{\mathbf{y}}\mp i\hat{\mathbf{x}}).
\end{eqnarray}
Using the definition of the energy flux density $\mathbf{S}=\mathbf{E}\times\mathbf{H}$~\cite{Barnett2016a}, we calculate
\begin{eqnarray}\label{eq:energyflux}
\mathbf{S}^{\pm}=\Re(\mathbf{E}^{\pm}\times(\mathbf{H}^{\pm})^*)=2|E_0|^2\sqrt{\frac{\epsilon}{\mu}}\hat{\mathbf{z}}.
\end{eqnarray}
As mentioned above, we extend the definition of the helicity flux in vacuum (\ref{eq:fluxvacuum}) to that in an isotropic medium by the replacement $\epsilon_0\rightarrow\epsilon$ and $\mu_0\rightarrow\mu$~\cite{VanKruining2016}, to find
\begin{eqnarray}\label{eq:flux}
\mathbf{v}^{\pm}& = \Re\left[\frac{1}{2}\left(\sqrt{\frac{\epsilon}{\mu}}\mathbf{E}^{\pm}\times(\mathbf{A}^{\pm})^*+\sqrt{\frac{\mu}{\epsilon}}\mathbf{H}^{\pm}\times(\mathbf{C}^{\pm})^*\right)\right]\nonumber\\
& = \pm\frac{1}{\omega}2|E_0|^2\sqrt{\frac{\epsilon}{\mu}}\hat{\mathbf{z}}.
\end{eqnarray}
The ratio of the energy flux density to helicity flux density is therefore given by
\begin{eqnarray}
\frac{\mathbf{v}^{\pm} \cdot \hat{\mathbf{z}}}{\mathbf{S}^{\pm}\cdot \hat{\mathbf{z}}}=\pm\frac{1}{\omega},
\end{eqnarray}
in agreement with vacuum value (\ref{eq:fluxratios}).

Second, we consider the helicity and energy densities. For almost all materials the chirality parameter, $\beta$, is small, and it is sufficient to work to first order in $\beta$. We do so for simplicity here, and consider higher powers of $\beta$ in the next section.  {From the energy density in a chiral medium~(\ref{eq:energydensity}), we calculate the energy density for right- and left-handed circularly polarised waves}
\begin{eqnarray}\label{eq:energydensitycalculated}
 w^{\pm}_{1} =2|E_0|^2\left(\epsilon(1\pm\beta k)\pm\epsilon\mu\beta\sqrt{\frac{\epsilon}{\mu}}\omega \right).
\end{eqnarray}
To obtain the ``na\"ive'' extension of the helicity density, we simply replace the values of permeability and permittivity in  (\ref{eq:helicityvacuum}) with those in the chiral medium to calculate
\begin{eqnarray}\label{eq:helicitydensitycalculated}
h^{\pm}&=\Re\left[\frac{1}{2}\left(\sqrt{\frac{\epsilon}{\mu}}(\mathbf{A}^{\pm})^*\cdot\mathbf{B}^{\pm}-\sqrt{\frac{\mu}{\epsilon}}(\mathbf{C}^{\pm})^*\cdot\mathbf{D}^{\pm}\right)\right]\nonumber\\
&=\pm \frac{1}{\omega}2|E_0|^2\epsilon(1\pm\beta k). 
\end{eqnarray}
Comparing the energy and helicity densities (\ref{eq:energydensitycalculated}) and (\ref{eq:helicitydensitycalculated}), it is clear that if we wish for the ratio of helicity density to energy density to be maintained inside the chiral medium, the helicity density is missing a term proportional to the chirality parameter $\beta$. We can recover the ratio ${{h}^{\pm}}/{{w}^{\pm}}=\pm 1/\omega$  by adding 
\begin{eqnarray}
\Re\left[\frac{1}{2}\sqrt{\epsilon\mu}\beta\left(\epsilon|\mathbf{E}|^2+\mu|\mathbf{H}|^2\right)\right]=2|E_0|^2\epsilon\mu\beta\sqrt{\frac{\epsilon}{\mu}}
\end{eqnarray}
 to  (\ref{eq:helicitydensitycalculated}), which motivates the following definition of helicity density  in a chiral medium to  $\mathcal{O}(\beta)$:
\begin{eqnarray}\label{eq:helicitybeta}
h_{1}&=\frac{1}{2}\left(\sqrt{\frac{\epsilon}{\mu}}\mathbf{A}\cdot\mathbf{B}-\sqrt{\frac{\mu}{\epsilon}}\mathbf{C}\cdot\mathbf{D}+\sqrt{\epsilon\mu}\beta\left(\epsilon|\mathbf{E}|^2+\mu|\mathbf{H}|^2\right)\right)\nonumber\\
&
\equiv\,h+\sqrt{\epsilon\mu}\beta\,w,
\end{eqnarray}
where $h$ is the  usual form helicity density in linear media (\ref{eq:helicityvacuum}), with appropriate replacement of the vacuum permeability and permittivity values. This modified form of the helicity densityis the central result of this paper.

\section{The helicity continuity equation in a dual-symmetric chiral medium}
We now need to establish  that the definition of helicity density (\ref{eq:helicitybeta}), motivated by consideration of circularly polarised plane waves, is appropriate for general fields inside a chiral medium. If we continue to consider a dual-symmetric medium, then helicity should be locally conserved, and we should be able to express this in a local continuity equation analogous to the vacuum continuity equation (\ref{eq:continuity}). We can form such an equation from our helicity density (\ref{eq:helicitybeta}). Taking the time derivative  and rearranging the expression results in
\begin{eqnarray}\label{eq:ongoing}
&\partial_t h_{1}+\frac{1}{2}\left(\sqrt{\frac{\epsilon}{\mu}}\mathbf{\mbox{\boldmath$\nabla$}}\cdot(\mathbf{E}\times\mathbf{A})+\sqrt{\frac{\mu}{\epsilon}}\mathbf{\mbox{\boldmath$\nabla$}}\cdot(\mathbf{H}\times\mathbf{C})\right)=
-\sqrt{\frac{\epsilon}{\mu}}\mathbf{E}\cdot\mathbf{B}\nonumber\\
&+\sqrt{\frac{\mu}{\epsilon}}\mathbf{H}\cdot\mathbf{D}+\sqrt{\epsilon\mu}\beta\left(\epsilon\mathbf{E}\cdot\dot{\mathbf{E}}+\mu\mathbf{H}\cdot\dot{\mathbf{H}}\right).
\end{eqnarray}
We  use the DBF relations in the time domain, $\mathbf{D}=\epsilon(\mathbf{E}-\beta\dot{\mathbf{B}})$, $\mathbf{B}=\mu(\mathbf{H}+\beta\dot{\mathbf{D}})$~\cite{Barnett2016a}, and retain only terms up to $\mathcal{O}(\beta)$, to write the right-hand side of (\ref{eq:ongoing}) as
\begin{eqnarray}\label{eq:23}
&-\sqrt{\frac{\epsilon}{\mu}}\mu\beta\mathbf{E}\cdot\dot{\mathbf{D}}-\sqrt{\frac{\mu}{\epsilon}}\epsilon\beta\mathbf{H}\cdot\dot{\mathbf{B}}\nonumber\\
&+\sqrt{\epsilon\mu}\beta\left(\epsilon\mathbf{E}\cdot\dot{\mathbf{E}}+\mu\mathbf{H}\cdot\dot{\mathbf{H}}\right)=0.
\end{eqnarray}
If $\mathbf{\mbox{\boldmath$\nabla$}}(\epsilon/\mu)=0$ throughout the material,  the left-hand side of (\ref{eq:ongoing}) then becomes
\begin{eqnarray}\label{eq:continuitymedium}
\partial_t h_{1}+\mathbf{\mbox{\boldmath$\nabla$}}\cdot\mathbf{v}=0,
\end{eqnarray}
showing that the form of the helicity density in (\ref{eq:helicitybeta}) is required for the local conservation of helicity within a dual-symmetric chiral medium. 

It is worth mentioning that chiral media are also examined in~\cite{VanKruining2016}, where it is shown that the conservation of helicity  of the form $h$  (\ref{eq:helicityvacuum})  for a chiral medium can be similarly expressed by a continuity equation. However, this result is derived to first order in the chirality parameter for the specific case of monochromatic fields with time dependence $\exp(-i\omega t)$ where the DBF relations reduce to $\mathbf{D}=\epsilon(\mathbf{E}+i\mu\beta\mathbf{H})$ and $\mathbf{B}=\mu(\mathbf{H}-i\epsilon\beta\mathbf{E})$. In this case, the real part of the  time derivative of the chiral contribution to the helicity density, $\sqrt{\epsilon\mu}\beta(\epsilon\mathbf{E}\cdot\dot{\mathbf{E}}+\mu\mathbf{H}\cdot\dot{\mathbf{H}})$, is zero, so that helicity conservation  indeed holds. For fields of a more general form, however, we stress that the helicity density of the form (\ref{eq:helicitybeta}) should be used for local  conservation within chiral media.  

\section{Higher powers of the chirality parameter}

Throughout this article so far, we have worked to first order in the chirality parameter $\beta$. We now consider   $\mathcal{O}(\beta^2)$ terms in the helicity density, by retaining terms $\mathcal{O}(\beta)$ in the energy density, to write
\begin{eqnarray}\label{eq:h2beta}
h_{2}=\frac{1}{2}\left(\sqrt{\frac{\epsilon}{\mu}}\mathbf{A}\cdot\mathbf{B}-\sqrt{\frac{\mu}{\epsilon}}\mathbf{C}\cdot\mathbf{D}\right)+\sqrt{\epsilon\mu}\beta w_{1},
\end{eqnarray}
where $w_{1}$ is the energy density to $\mathcal{O}(\beta)$, as given in equation (\ref{eq:energydensity}). Repeating the above treatment, it is possible to show $\partial_t h_{2}+\mathbf{\mbox{\boldmath$\nabla$}}\cdot\mathbf{v}=0$. 
 The subscript on $h_2$  indicates   that only terms $\mathcal{O}(\beta^2)$ are retained to produce the continuity equation $\partial_t h_{2}+\mathbf{\mbox{\boldmath$\nabla$}}\cdot\mathbf{v}=0$, in the same way that only terms $\mathcal{O}(\beta)$ are retained on the right-hand side of (\ref{eq:ongoing}) to produce (\ref{eq:23}). That is, the truncation to $\mathcal{O}(\beta^2)$ is performed only after the time derivative is taken.
If we were working to $\mathcal{O}(\beta^3)$, we would retain terms of $\mathcal{O}(\beta^3)$ in the time derivative of $h+\sqrt{\epsilon\mu}\beta w_{2}$, and so on. Indeed, it is straightforward to show using (\ref{eq:DBF}) that the  expression for  the energy density ~\cite{Fedorov1976,Proskurin2017}
\begin{eqnarray}\label{eq:exactenergy}
w_{\beta}=\frac{1}{2}\left(\frac{1}{\epsilon}\mathbf{D}\cdot\mathbf{D}+\frac{1}{\mu}\mathbf{B}\cdot\mathbf{B}\right)
\end{eqnarray}
satisfies the condition for local energy conservation exactly. Making the replacement $\epsilon_0\rightarrow\epsilon$ and $\mu_0\rightarrow\mu$ in (\ref{eq:helicityvacuum})   and (\ref{eq:flux}) and again using the DBF constitutive relations leads to 
\begin{eqnarray}\label{eq:hconservation}
\partial_th+\mbox{\boldmath$\nabla$}\cdot\mathbf{v}=\sqrt{\epsilon\mu}\beta\mbox{\boldmath$\nabla$}\cdot(\mathbf{E}\times\mathbf{H}).
\end{eqnarray}
Recognising the divergence term on the right-hand side of (\ref{eq:hconservation}) as Poynting's vector, the conservation of helicity can then be expressed  succinctly  as
\begin{eqnarray}\label{eq:26}
\partial_t\left(h+\sqrt{\epsilon\mu}\beta w_{\beta}\right)+\mbox{\boldmath$\nabla$}\cdot\mathbf{v}&=\sqrt{\epsilon\mu}\beta\mbox{\boldmath$\nabla$}\cdot\mathbf{S}-\sqrt{\epsilon\mu}\beta(\mbox{\boldmath$\nabla$}\cdot\mathbf{S})\nonumber\\
&=0.
\end{eqnarray}
Thus,
\begin{eqnarray}\label{eq:hbetaall}
h_{\beta}=\frac{1}{2}\left(\sqrt{\frac{\epsilon}{\mu}}\mathbf{A}\cdot\mathbf{B}-\sqrt{\frac{\mu}{\epsilon}}\mathbf{C}\cdot\mathbf{D}\right)+\sqrt{\epsilon\mu}\beta w_{\beta}
\end{eqnarray}
is an exact expression for  the electromagnetic helicity in a chiral medium. It is worthwhile mentioning that expanding the energy density (\ref{eq:exactenergy}) using the DBF constitutive relations in the time domain leads to an infinite series of terms of increasing order in $\beta$. The helicity density to $\mathcal{O}(\beta^{n+1})$ then takes the form 
\begin{eqnarray}\label{eq:hnbeta}
h_{n+1}=\frac{1}{2}\left(\sqrt{\frac{\epsilon}{\mu}}\mathbf{A}\cdot\mathbf{B}-\sqrt{\frac{\mu}{\epsilon}}\mathbf{C}\cdot\mathbf{D}\right)+\sqrt{\epsilon\mu}\beta w_{n}\,\forall n,
\end{eqnarray}
where the subscript of $w_n$ indicates that only terms  up to $\mathcal{O}(\beta^n)$ are retained in the expansion of the energy density in the time domain. 

 As the energy and helicity fluxes do not change with increasing orders in $\beta$, it follows that the speed at which the densities propagate within the medium is dependent upon $\beta$. This is confirmed by  calculation  of the group velocity of the right- and left-handed circularly polarised waves within the chiral medium.
 From Maxwell's equations and the constitutive  relations (\ref{eq:DBF}), we can obtain the wave equation within a chiral medium:
\begin{eqnarray}
\mathbf{\mbox{\boldmath$\nabla$}}^2\mathbf{H}=\epsilon\mu\ddot{\mathbf{H}}+2\epsilon\mu\beta\mathbf{\mbox{\boldmath$\nabla$}}\times\ddot{\mathbf{H}}-\epsilon\mu\beta^2\mathbf{\mbox{\boldmath$\nabla$}}^2\ddot{\mathbf{H}}.
\end{eqnarray}
Inserting $\mathbf{H}^{\pm}$ from (\ref{eq:EandH}) leads to the dispersion relation
$\omega^{\pm}={k}/{\sqrt{\epsilon\mu}(1\pm\beta k)}$, so that the  group velocity is  given by
\begin{eqnarray}\label{eq:groupv}
{v}_g^{\pm}=\frac{d \omega^{\pm}}{d k}=\frac{1}{\sqrt{\epsilon\mu}(1\pm \beta k)^2}.
\end{eqnarray}

The expression for  $h_{\beta}$ in (\ref{eq:hbetaall}) for right- and left-handed circular polarised plane waves  leads to a  speed of propagation of the helicity density of the wave in agreement with  (\ref{eq:groupv}). Considering again $\mathbf{E}^{\pm}$ and $\mathbf{H}^{\pm}$ in (\ref{eq:EandH}), we calculate
\begin{eqnarray}\label{eq:hbetaongoing}
 h_{\beta}^{\pm} &=\pm\frac{1}{\omega}2|E_0|^2\epsilon(1\pm\beta k)+2\beta\sqrt{\epsilon\mu}(1\pm\beta k)^2.
\end{eqnarray}
Using the dispersion relation, this can be rewritten
\begin{eqnarray}\label{eq:hbetaongoing2}
 h_{\beta}^{\pm} &=\pm\frac{1}{\omega}2\epsilon|E_0|^2(1\pm\beta k)^2, 
\end{eqnarray}
so that inserting the flux density of the fields from   (\ref{eq:flux}) leads to : 
\begin{eqnarray}
\frac{\mathbf{v}^{\pm}\cdot\hat{\mathbf{z}}}{ h_{\beta}^{\pm}}=\frac{1}{\sqrt{\epsilon\mu}(1\pm \beta k)^2}.
\end{eqnarray}
An analogous result holds for the propagation speed of the energy density, $\mathbf{S}^{\pm}\cdot\hat{\mathbf{z}}/w_{\beta}^{\pm}=v_g^{\pm}$, from which it follows that the ratio
\begin{eqnarray}
\frac{h_{\beta}^{\pm}}{w_{\beta}^{\pm}}=\pm\frac{1}{\omega}
\end{eqnarray}
holds for the circularly  polarised plane waves inside the chiral medium.

\section*{Conclusion} 
The  dual symmetry of the free-space Maxwell equations generated by the conservation of helicity invariably underpins much of classical electromagnetism. That this symmetry also exists  within some media is  a surprising result  which provides us with an elegant way by which to probe the chiral response of matter~\cite{Fernandez-Corbaton2013a,VanKruining2016,Alpeggiani2018}, and ultimately to relate this to the response of individual microscopic sources of helicity~\cite{Nienhuis2016,Crimin2019,Fernandez-Corbaton2015,Fernandez-Corbaton2016}. 

In this paper, we have used the dual symmetry of the macroscopic Maxwell equations under certain conditions to  review the resulting conservation of helicity. Using the DBF constitutive relations (\ref{eq:DBF}), we  discussed that the condition for dual symmetry imposes no restriction upon  a  chirality parameter $\beta\in\Re$ of a chiral medium, and so it follows  that helicity is conserved within such media where there is no gradient in $\epsilon/\mu$.  We examined a simple case where we know that helicity  must be conserved: the propagation of right- and left-handed circular polarised plane waves  traversing a vacuum-chiral interface where $\epsilon/\mu=\epsilon_0/\mu_0$, and used this to motivate an expression for the helicity density of electromagnetic fields within a chiral medium. We extended the result to include higher orders in the chirality parameter, and tested the resulting  expression by various means:  it allows us to express the conservation of helicity in a local continuity equation, it  produces the correct group velocity for the right- and left-handed  circularly polarised plane waves, and it  leads to the correct ratio of helicity density to energy density for these waves inside the medium. The local   conservation of helicity in a chiral medium then follows as a consequence of the conservation of energy,  as follows from (\ref{eq:26}).  

 \section*{Acknowledgements}
 We would like to thank Igor Proskurin for providing us with a copy of the  reference~\cite{Fedorov1976}  and  for  his helpful comments on the exact form of the energy density in a chiral medium. We acknowledge funding from The Royal Society under grant numbers RP/EA/180010 and RP/150122, and the Engineering and Physical Sciences Research Council  under grant number EP/N509668/1.

\end{document}